\documentstyle[12pt,epsfig,amssymb]{article}

\catcode`\@=11
\long\def\@makefntext#1{
\protect\noindent \hbox to 3.2pt {\hskip-.9pt  
$^{{\ninerm\@thefnmark}}$\hfil}#1\hfill}                

\def\@makefnmark{\hbox to 0pt{$^{\@thefnmark}$\hss}}  
        
\def\ps@myheadings{\let\@mkboth\@gobbletwo
\def\@oddhead{\hbox{}
\rightmark\hfil\ninerm\thepage}   
\def\@oddfoot{}\def\@evenhead{\ninerm\thepage\hfil
\leftmark\hbox{}}\def\@evenfoot{}
\def\sectionmark##1{}\def\subsectionmark##1{}}

\renewcommand{\thefootnote}{\fnsymbol{footnote}}

\newcounter{sectionc}\newcounter{subsectionc}\newcounter{subsubsectionc}
\renewcommand{\section}[1] {\vspace*{0.6cm}\addtocounter{sectionc}{1} 
\setcounter{subsectionc}{0}\setcounter{subsubsectionc}{0}\noindent 
        {\normalsize\bf\thesectionc. #1}\par\vspace*{0.4cm}}
\renewcommand{\subsection}[1] {\vspace*{0.6cm}\addtocounter{subsectionc}{1} 
        \setcounter{subsubsectionc}{0}\noindent 
        {\normalsize\it\thesectionc.\thesubsectionc. #1}\par\vspace*{0.4cm}}
\renewcommand{\subsubsection}[1]
{\vspace*{0.6cm}\addtocounter{subsubsectionc}{1}
        \noindent {\normalsize\rm\thesectionc.\thesubsectionc.\thesubsubsectionc. 
        #1}\par\vspace*{0.4cm}}

\newcounter{appendixc}
\newcounter{subappendixc}[appendixc]
\newcounter{subsubappendixc}[subappendixc]

\renewcommand{\appendix}[1] {\vspace*{0.6cm}
        \refstepcounter{appendixc}
        \setcounter{figure}{0}
        \setcounter{table}{0}
        \setcounter{equation}{0}
        \renewcommand{\thefigure}{\Alph{appendixc}.\arabic{figure}}
        \renewcommand{\thetable}{\Alph{appendixc}.\arabic{table}}
        \renewcommand{\theappendixc}{\Alph{appendixc}}
        \renewcommand{\theequation}{\Alph{appendixc}.\arabic{equation}}
        \noindent{\bf Appendix \theappendixc #1}\par\vspace*{0.4cm}}

\def\abstracts#1{{
        \centering{\begin{minipage}{12.2truecm}\footnotesize\baselineskip=12pt\noindent
        \centerline{\footnotesize ABSTRACT}\vspace*{0.3cm}
        \parindent=0pt #1
        \end{minipage}}\par}} 

\renewenvironment{thebibliography}[1]
        {\begin{list}{\arabic{enumi}.}
        {\usecounter{enumi}\setlength{\parsep}{0pt}
\setlength{\leftmargin 1.25cm}{\rightmargin 0pt}
         \setlength{\itemsep}{0pt} \settowidth
        {\labelwidth}{#1.}\sloppy}}{\end{list}}

\newcounter{itemlistc}
\newcounter{romanlistc}
\newcounter{alphlistc}
\newcounter{arabiclistc}

\newcommand{\fcaption}[1]{
        \refstepcounter{figure}
        \setbox\@tempboxa = \hbox{\footnotesize Fig.~\thefigure. #1}
        \ifdim \wd\@tempboxa > 6in
           {\begin{center}
        \parbox{6in}{\footnotesize\baselineskip=12pt Fig.~\thefigure. #1}
            \end{center}}
        \else
             {\begin{center}
             {\footnotesize Fig.~\thefigure. #1}
              \end{center}}
        \fi}

\newcommand{\tcaption}[1]{
        \refstepcounter{table}
        \setbox\@tempboxa = \hbox{\footnotesize Table~\thetable. #1}
        \ifdim \wd\@tempboxa > 6in
           {\begin{center}
        \parbox{6in}{\footnotesize\baselineskip=12pt Table~\thetable. #1}
            \end{center}}
        \else
             {\begin{center}
             {\footnotesize Table~\thetable. #1}
              \end{center}}
        \fi}

\def\@citex[#1]#2{\if@filesw\immediate\write\@auxout
        {\string\citation{#2}}\fi
\def\@citea{}\@cite{\@for\@citeb:=#2\do
        {\@citea\def\@citea{,}\@ifundefined
        {b@\@citeb}{{\bf ?}\@warning
        {Citation `\@citeb' on page \thepage \space undefined}}
        {\csname b@\@citeb\endcsname}}}{#1}}

\newif\if@cghi
\def\cite{\@cghitrue\@ifnextchar [{\@tempswatrue
        \@citex}{\@tempswafalse\@citex[]}}
\def\citelow{\@cghifalse\@ifnextchar [{\@tempswatrue
        \@citex}{\@tempswafalse\@citex[]}}
\def\@cite#1#2{{$\null^{#1}$\if@tempswa\typeout
        {IJCGA warning: optional citation argument 
        ignored: `#2'} \fi}}

 1
 1
 1

\font\ninerm=cmr9

\begin{document}
\setlength{\baselineskip}{0.75cm}
\begin{titlepage}
\begin{flushright}
\begin{tabular}{l}
CERN-TH/97-189 \\
DO-TH 97/17 \\
\end{tabular}
\end{flushright}
\vspace*{0.8cm}
\begin{center}
\Large{
{\bf Some Aspects of \\}
\vspace*{0.5cm}
{\bf Polarized $ep$ Scattering at HERA}\footnote{Invited talk presented
by W.\ Vogelsang at the Ringberg Workshop on New Trends in HERA Physics,
Ringberg Castle, Germany, 25-30 May 1997.} \\}
\vskip 2cm
{\large Marco Stratmann}  \\
\vspace*{0.3cm}
\normalsize
Institut f\"{u}r Physik, Universit\"{a}t Dortmund, \\ 
D-44221 Dortmund, Germany \\
\vspace*{0.3cm}
and \\
\vspace*{0.5cm}
{\large Werner Vogelsang}  \\
\vspace*{0.3cm}
\normalsize
Theory Division, CERN,\\
CH-1211 Geneva 23, Switzerland \\
\end{center}
\vskip 1.8cm
\begin{center}
{\bf Abstract} \\
\end{center}
We study photoproduction reactions
in a polarized $ep$ collider mode of HERA at $\sqrt{s}=298$ GeV.
We examine the sensitivity of the cross sections and their asymmetries 
to the proton's polarized gluon distribution and to the completely unknown 
parton distributions of longitudinally polarized photons. \\ \\ \\
CERN-TH/97-189 \\
DO-TH 97/17 \\
July 1997 
\end{titlepage}

\parindent=1.5pc
\baselineskip=15pt

\centerline{\normalsize\bf SOME ASPECTS OF POLARIZED $ep$ SCATTERING AT HERA}

\vspace*{0.6cm}
\centerline{\footnotesize M. Stratmann}
\baselineskip=13pt
\centerline{\footnotesize\it Institut f\"{u}r Physik, Universit\"{a}t 
Dortmund, D-44221 Dortmund, Germany}
\vspace*{0.3cm}
\centerline{\footnotesize and}
\vspace*{0.3cm}
\centerline{\footnotesize W. Vogelsang}
\baselineskip=13pt
\centerline{\footnotesize\it Theory Division, CERN, CH-1211 Geneva, 
Switzerland}

\vspace*{0.6cm}
\abstracts{We study photoproduction reactions
in a polarized $ep$ collider mode of HERA at $\sqrt{s}=298$ GeV.
We examine the sensitivity of the cross sections and their asymmetries 
to the proton's polarized gluon distribution and to the completely unknown 
parton distributions of longitudinally polarized photons.}
 
\normalsize\baselineskip=15pt
\setcounter{footnote}{0}
\renewcommand{\thefootnote}{\alph{footnote}}
\section{Introduction}
Among the various conceivable options for future HERA upgrades is the
idea to longitudinally polarize its proton beam~\cite{sch} which, 
when combined with the already operative longitudinally polarized electron 
(positron) beam, results in a polarized version of the usual HERA collider 
with $\sqrt{S}\approx 300$ GeV. A typical conservative value for the 
integrated luminosity in this case should be 100 $\mbox{pb}^{-1}$.

HERA has already been very successful in pinning down 
the proton's unpolarized gluon distribution $g(x,Q^2)$. Several processes 
have been studied which have contributions from $g(x,Q^2)$ already in the 
lowest order, such as (di)jet, inclusive hadron, and heavy flavor production. 
Since events at HERA are concentrated in the region $Q^2 \rightarrow 0$, the 
processes have first and most accurately been studied in 
photoproduction$^{2\mbox{-}9}$. As is well-known, in this case the 
(quasi-real) photon will not only interact in a direct (`point-like') way, 
but can also be resolved into its hadronic structure. HERA photoproduction 
experiments like$^{2\mbox{-}9}$
have not merely established evidence for the existence of such a resolved
contribution, but have also been precise enough to improve our knowledge
about the parton distributions, $f^{\gamma}$, of the photon.

Given the success of such unpolarized photoproduction experiments at HERA, 
it seems most promising~\cite{wir} to closely examine the same processes for 
the situation with longitudinally polarized beams with regard to their 
sensitivity to the proton's polarized gluon distribution $\Delta g$, which is
still one of the most interesting, but least known, quantities in
`spin-physics'. Recent next-to-leading (NLO) studies of polarized
DIS~\cite{grsv,gs} show that the $x$-shape of $\Delta g$ seems to be 
hardly constrained at all by the present DIS data, even though
a tendency towards a sizeable positive {\em total} gluon polarization,
$\int_0^1 \Delta g(x,Q^2=4 \; \mbox{GeV}^2) dx \gtrsim 1$, was found
\cite{grsv,bfr,gs}. Furthermore, polarized photoproduction experiments 
may in principle allow to not only determine the parton, in
particular gluon, content of the polarized {\em proton}, but also 
that of the longitudinally polarized {\em photon} which is completely
unknown so far. Since a measurement of, e.g., the photon's spin-dependent
structure function $g_1^{\gamma}$ in polarized $e^+ e^-$ collisions is not
planned in the near future, HERA could play a unique role here, even if it
should only succeed in establishing the very {\em existence} of a resolved
contribution to polarized photon-proton reactions.

Our contribution, part of which is taken from~\cite{wir}, is organized 
as follows: In the next section we collect the necessary ingredients for our 
calculations. In section 3 we will discuss various conceivable 
photoproduction reactions, namely (di)jet, inclusive hadron, and open-charm
production.

\section{Polarized Parton Distributions of the Proton and the Photon}
Even though NLO analyses of polarized DIS which take into account all or 
most data sets have been published recently~\cite{grsv,bfr,gs},
we have to stick to LO calculations throughout this work since the NLO 
corrections to, e.g., polarized jet production are not yet known. This 
implies use of LO parton distributions, which have also been provided in the 
studies~\cite{grsv,gs}. Both papers give various LO sets which mainly differ 
in the $x$-shape of the polarized gluon distribution. We will choose the LO 
`valence' set of the `radiative parton model analysis'~\cite{grsv}, which 
corresponds to the best-fit result of that paper, along with two other sets 
of~\cite{grsv} which are based on either assuming $\Delta g (x,\mu^2) = 
g(x,\mu^2)$ or $\Delta g(x,\mu^2)=0$ at the low input scale $\mu$ of 
\cite{grsv}, where $g(x,\mu^2)$ is the unpolarized LO GRV~\cite{grv} input 
gluon distribution. These two sets will be called `$\Delta g=g$ input' and 
`$\Delta g=0$ input' scenarios, respectively. The gluon of set C of~\cite{gs}
is qualitatively different since it has a substantial negative polarization 
at large $x$. We will therefore also use this set in our calculations.
For illustration, we show in Fig.~1 the gluon distributions  
of the four different sets of parton distributions we will use, taking a 
typical scale $Q^2=10$ GeV$^2$. Keeping in mind that all four LO sets provide 
very good descriptions of the present polarized DIS data, 
it becomes obvious that the data indeed do not seem to be able to 
significantly constrain the $x$-shape of $\Delta g(x,Q^2)$.
\begin{figure}[ht]
\begin{center}
\vspace*{-1cm}
\hspace*{0.5cm}
\epsfig{file=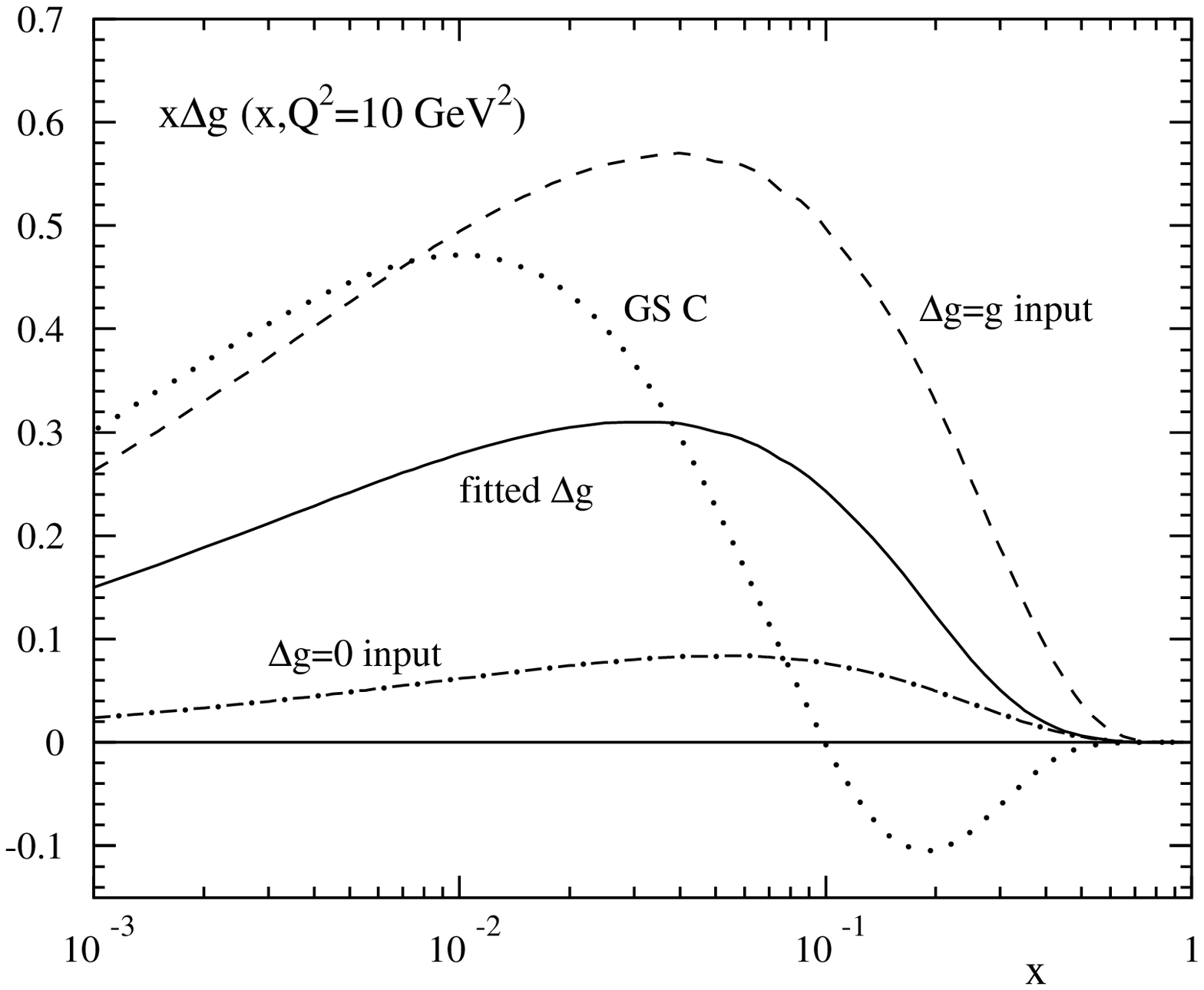,width=11cm}
\vspace*{-0.8cm}
\fcaption{\it Gluon distributions at $Q^2=10\,{\it{GeV}}^{\:2}$ 
of the four LO sets of
polarized parton distributions used in this paper. The dotted line refers
to set C of~\cite{gs}, whereas the other distributions are taken 
from~\cite{grsv} as described in the text.}
\vspace*{-0.3cm}
\end{center}
\end{figure}

In the case of photoproduction the electron just
serves as a source of quasi-real photons which are radiated according 
to the Weizs\"{a}cker-Williams spectrum. The photons can then 
interact either directly or via their partonic structure (`resolved' 
contribution). In the case of longitudinally polarized electrons, the 
resulting photon will be longitudinally (more precisely, circularly)
polarized and, in the resolved case, the {\em polarized} parton 
distributions of the photon, $\Delta f^{\gamma}(x,Q^2)$, enter the 
calculations. Thus one can define the effective polarized parton densities 
at the scale $M$ 
\nopagebreak[4]
in the longitudinally polarized electron via\footnote{We 
include here the additional definition 
$\Delta f^{\gamma} (x_{\gamma},M^2) \equiv \delta (1-x_{\gamma})$ for the 
direct (`unresolved') case.}
\pagebreak[4]
\begin{equation}  \label{elec}
\Delta f^e (x_e,M^2) = \int_{x_e}^1 \frac{dy}{y} \Delta P_{\gamma/e} (y)
\Delta f^{\gamma} (x_{\gamma}=\frac{x_e}{y},M^2) \; 
\end{equation}
($f=q,g$) where $\Delta P_{\gamma/e}$ is the polarized Weizs\"{a}cker-Williams
spectrum for which we will use 
\begin{equation}  \label{weiz}
\Delta P_{\gamma/e} (y) = \frac{\alpha_{em}}{2\pi} \left[ 
\frac{1-(1-y)^2}{y} \right] \ln \frac{Q^2_{max} (1-y)}{m_e^2 y^2} \; ,
\end{equation}
with the electron mass $m_e$. For the time being, it seems most 
sensible to follow as closely as possible the analyses successfully 
performed in the unpolarized case, which implies to introduce the same 
kinematical cuts. As in~\cite{jet1ph,jet2ph,chph,kramer} we will use
an upper cut\footnote{In H1 analyses of HERA photoproduction
data~\cite{jet1h1,jet2h1,chh1}
the cut $Q^2_{max}=0.01$ GeV$^2$ is used along with slightly
different $y$-cuts as compared to the corresponding 
ZEUS measurements~\cite{jet1ph,jet2ph,chph}, which leads to
smaller rates.} 
$Q^2_{max}=4$ GeV$^2$, and the $y$-cuts $0.2 \leq y \leq 0.85$ 
(for single-jet~\cite{jet1ph} and charm production) 
and $0.2 \leq y \leq 0.8$ (for dijet~\cite{jet2ph} and inclusive hadron 
production) will be imposed. 

The polarized photon structure functions $\Delta f^{\gamma} (x_{\gamma},M^2)$ 
in (\ref{elec}) are completely unmeasured so far, such that 
models for them have to be invoked. To obtain a realistic estimate for the 
theoretical uncertainties in the polarized photonic parton densities
two very different scenarios were considered in~\cite{gvg} assuming 
`maximal' ($\Delta f^{\gamma}(x,\mu^2)=f^{\gamma}(x,\mu^2)$) or `minimal' 
($\Delta f^{\gamma}(x,\mu^2)=0$) saturation of the fundamental positivity 
constraints $|\Delta f^{\gamma}(x,\mu^2)| \leq f^{\gamma}(x,\mu^2)$ at the
input scale $\mu$ for the QCD evolution. Here $\mu$ and the unpolarized 
photon structure functions $f^{\gamma}(x,\mu^2)$ were adopted from the 
phenomenologically successful radiative parton model predictions in 
\cite{grvg}. The results of these two extreme approaches are presented in 
Fig.~2 in terms of the photonic parton asymmetries $A_f^{\gamma} \equiv 
\Delta f^{\gamma}/f^{\gamma}$, evolved to $Q^2=30$ GeV$^2$ in LO. An ideal 
aim of measurements in a polarized collider mode of HERA would of course be 
to determine the $\Delta f^{\gamma}$ and to see which ansatz is more 
realistic. The sets presented in Fig.~2, which we 
will use in what follows, should in any case be sufficient to study 
the sensitivity of the various cross sections to the $\Delta f^{\gamma}$,
but also to see in how far they influence a determination of $\Delta g$.
\begin{figure}[ht]
\begin{center}
\vspace*{-1cm}
\hspace*{0.4cm}
\epsfig{file=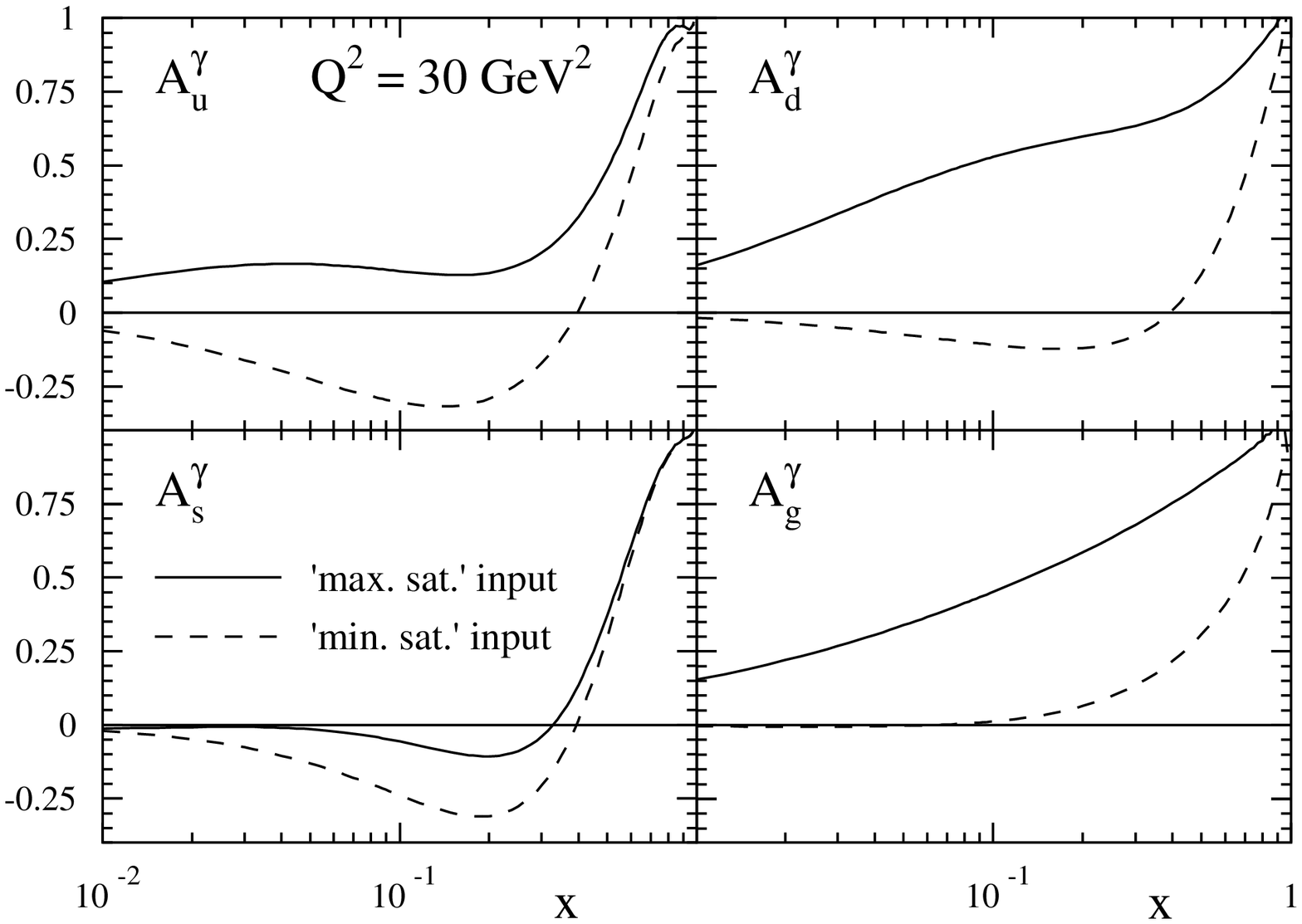,width=12cm}
\vspace*{-0.4cm}
\fcaption{\it Photonic LO parton asymmetries 
$A_f^{\gamma}\equiv \Delta f^{\gamma}/f^{\gamma}$ at 
$Q^2=30\,{\it{GeV}}^{\,2}$ 
for the two scenarios considered in~\cite{gvg} (see text). The 
unpolarized LO photonic parton distributions were taken from~\cite{grvg}.}
\vspace*{-0.3cm}
\end{center}
\end{figure}

We finally note that in what follows a polarized cross section will always
be defined as
\begin{equation} 
\Delta \sigma \equiv \frac{1}{2} \left( \sigma (++)-\sigma (+-) \right) \; ,
\end{equation}
the signs denoting the helicities of the scattering particles. The
corresponding unpolarized cross section is given by taking the sum 
instead, and the cross section asymmetry is $A\equiv \Delta \sigma/\sigma$.
Whenever calculating an asymmetry $A$, we will 
use the LO GRV parton distributions for the proton~\cite{grv} and the 
photon~\cite{grvg} to calculate the unpolarized cross section.
For consistency, we will employ the LO expression for the strong coupling 
$\alpha_s$ with~\cite{grsv,gs,gvg} $\Lambda_{QCD}^{(f=4)}=200$ MeV for 
four active flavors.

\pagebreak[4]
\section{Photoproduction Reactions at Polarized HERA}
The generic LO cross section formula for the photoproduction of a single jet 
with transverse momentum $p_T$ and cms-rapidity $\eta$ in polarized $ep$ 
collisions reads:
\begin{equation} \label{wqc}
\frac{d^2 \Delta \sigma}{dp_T d\eta} =  \sum_{f^e,f^p,c} 
\Delta f^e (x_e,M^2) \otimes \Delta f^p (x_p,M^2) \otimes 
\frac{d^2 \Delta \hat{\sigma}^{f_e f_p \rightarrow cd}}{dp_T d\eta} \; ,
\end{equation}
where $\otimes$ denotes a convolution and the sum is running over all 
properly symmetrized $2\rightarrow 2$ subprocesses for the direct 
($\gamma b\rightarrow cd$, $\Delta f^e (x_e,M^2) \equiv 
\Delta P_{\gamma/e}(x_e)$) and resolved ($ab\rightarrow cd$) cases. When only 
light flavors are involved, the corresponding differential helicity-dependent 
LO subprocess cross sections can be found in~\cite{bab}. In all following 
predictions we will deal with the charm contribution to the cross section 
by including charm only as a {\em{final}} state particle produced via the 
subprocesses $\gamma g \rightarrow c\bar{c}$ (for the direct part) and $gg 
\rightarrow c\bar{c}$, $q\bar{q} \rightarrow c\bar{c}$ (for the resolved 
part). For the values of $p_T$ considered it turns out that the finite 
charm mass 
can be safely neglected in these subprocess cross sections. In (\ref{wqc}), 
$\hat{s} \equiv x_e x_p s$ and $M$ is the factorization/renormalization scale 
for which we will use\footnote{The scale dependence of the theoretical LO
predictions for the spin asymmetries---which are the quantities relevant in 
experiments---turns out to be very weak, for a discussion see~\cite{wir}.}
$\; M=p_T$. The $\Delta f^p$ stand for the 
polarized parton distributions of the proton. Needless to say that we  
obtain the corresponding unpolarized LO jet cross section $d^2 \sigma/
dp_T d\eta$ by using LO unpolarized parton distributions and subprocess cross 
sections in (\ref{wqc}). 

It appears very promising~\cite{wir} to study the $\eta_{LAB}$-distribution of 
the cross section and the asymmetry, where $\eta_{LAB}$ is the laboratory
frame rapidity, related to $\eta$ via $\eta \equiv \eta_{cms} = \eta_{LAB} -
\frac{1}{2} \ln (E_p/E_e)$. As usual, $\eta_{LAB}$ is defined to be positive
in the proton forward direction. The crucial point is that for negative 
$\eta_{LAB}$ the main contributions are expected to come from the region 
$x_{\gamma} \rightarrow 1$ and thus mostly from the direct piece at 
$x_{\gamma}=1$. To investigate this, Fig.~3 shows our results for the 
single-inclusive jet cross section and its asymmetry vs. $\eta_{LAB}$ 
and integrated over $p_T>8$ GeV for the four sets of the polarized 
proton's parton distributions. For Figs.~3a,b we have used the `maximally' 
saturated set of polarized photonic parton densities, whereas Figs.~3c,d 
correspond to the `minimally' saturated one. Comparison of Figs.~3a,c or
3b,d shows that indeed the direct contribution clearly dominates for 
$\eta_{LAB} \leq -0.5$, where also differences between the polarized gluon 
distributions of the proton show up clearly. Furthermore, the cross sections 
are generally large in this region with asymmetries of a few percents. At 
positive $\eta_{LAB}$, we find that the cross section is 
dominated by the resolved contribution and is therefore sensitive to
the parton content of both the polarized proton {\em and} the photon.
This means that one can only learn something about the polarized photon 
structure functions if the polarized parton distributions 
\pagebreak[4]
\begin{figure}[t]
\begin{center}
\vspace*{-0.9cm}
\hspace*{0.5cm}
\epsfig{file=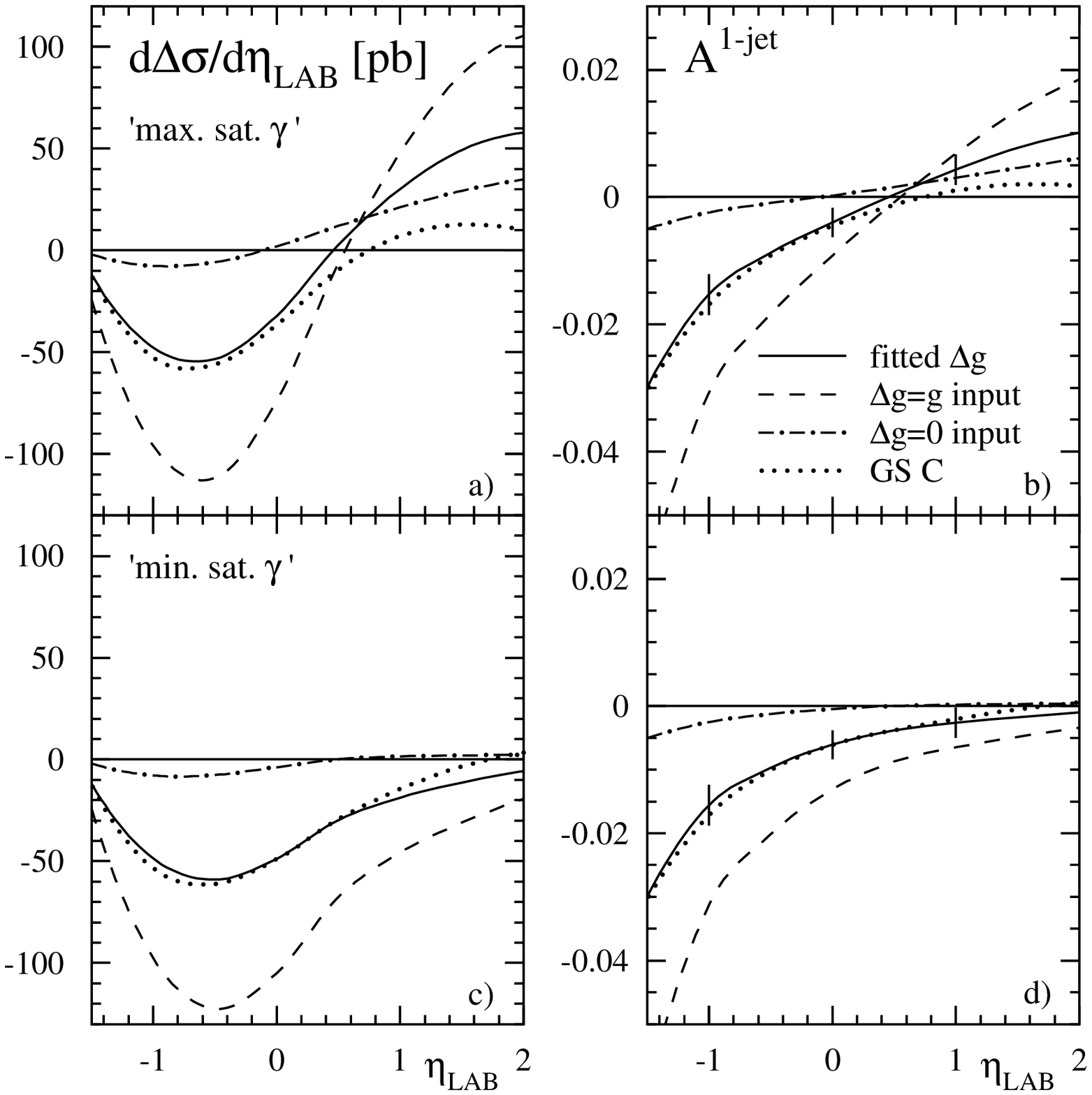,width=11.5cm}
\vspace*{-0.5cm}
\fcaption{\it {\bf a:} $\eta_{LAB}$-dependence of the polarized single-jet 
inclusive photoproduction cross section in $ep$-collisions at HERA, integrated
over $p_T > 8$ GeV. The renormalization/factorization scale was 
chosen to be $M=p_T$. The resolved contribution to the cross section has 
been calculated with the `maximally' saturated set of polarized photonic 
parton distributions. {\bf b:} Asymmetry corresponding to {\bf a}. The 
expected statistical errors have been calculated according to (\ref{aerr})
and as described in the text. {\bf c,d:} Same as {\bf a,b}, 
but for the `minimally' saturated set of polarized photonic parton 
distributions.}
\vspace*{-0.3cm}
\end{center}
\end{figure}
of the proton are 
already known to some accuracy or if an experimental distinction between 
resolved and direct contributions can be achieved. We note that the dominant 
contributions to the resolved part at large $\eta_{LAB}$ are driven by the 
polarized photonic {\em gluon} distribution $\Delta g^{\gamma}$. 
We have included in the asymmetry plots in Figs.~3b,d the expected 
statistical errors $\delta A$ at HERA which can be estimated from 
\begin{equation}  \label{aerr}
\delta A = \frac{1}{P_e P_p \sqrt{{\cal L} \sigma \epsilon}} \; ,
\end{equation}
where $P_e$, $P_p$ are the beam polarizations, ${\cal L}$ is the integrated 
luminosity and $\epsilon$ the jet detection efficiency, for which we assume
$P_e * P_p=0.5$, ${\cal L}=100$/pb and $\epsilon=1$. From the results it 
appears that a measurement of the proton's $\Delta g$ should be possible from 
single-jet events at negative rapidities where the contamination from the 
resolved contribution is minimal. 

In the unpolarized case, an experimental criterion for a distinction  
between direct and resolved contributions has been introduced~\cite{jeff} and 
used~\cite{jet2ph} in the case of dijet photoproduction at HERA. We will now 
adopt this criterion for the polarized case to see whether it would enable a
further access to $\Delta g$ and/or the polarized photon structure functions. 
The generic expression for the polarized cross section $d^3 \Delta \sigma/dp_T 
d\eta_1 d\eta_2$ for the photoproduction of two jets with laboratory system 
rapidities $\eta_1$, $\eta_2$ has a form analogous to (\ref{wqc}).
Here one has 
\begin{equation}
x_e \equiv \frac{p_T}{2 E_e} \left( e^{-\eta_1} + e^{-\eta_2} \right)\;\; , \;
x_p \equiv \frac{p_T}{2 E_p} \left( e^{\eta_1} + e^{\eta_2} \right) \; ,
\end{equation}
where $p_T$ is the transverse momentum of one of the two jets (which balance
each other in LO).
Following~\cite{jet2ph}, we will integrate over the cross section to obtain 
$d\Delta \sigma/d\bar{\eta}$, where $\bar{\eta} \equiv (\eta_1 + \eta_2)/2$.
Furthermore, we will apply the cuts~\cite{jet2ph}
$|\Delta \eta| \equiv |\eta_1-\eta_2| \leq 0.5 \; , \;\; 
p_T>6 \; \mbox{GeV}$.
The important point is that measurement of the jet rapidities allows 
for fully reconstructing the kinematics of the underlying hard subprocess
and thus for determining the variable~\cite{jet2ph}
\begin{equation}
x_{\gamma}^{OBS} = \frac{\sum_{jets} p_T^{jet} e^{-\eta^{jet}}}{2yE_e} \; ,
\end{equation} 
which in LO equals $x_{\gamma}=x_e/y$ with $y$ as before being the 
fraction of the electron's energy taken by the photon. Thus it becomes
possible to experimentally select events at large $x_{\gamma}$, 
$x_{\gamma} > 0.75$~\cite{jeff,jet2ph}, 
hereby extracting the {\em direct} contribution to 
the cross section with just a rather small contamination from resolved 
processes. Conversely, the events with $x_{\gamma}\leq 0.75$ will represent 
the resolved part of the cross section. This procedure should therefore be 
ideal to extract $\Delta g$ on the one hand, and examine the polarized 
photon structure functions on the other.

Fig.~4 shows the results~\cite{wir} for the direct part of the cross section 
according to the above selection criteria. The contributions from the 
resolved subprocesses have been included, using the `maximally' 
saturated set of polarized photonic parton densities. They turn out
to be non-negligible but, as expected, subdominant. More importantly,
due to the constraint $x_{\gamma}>0.75$ they are determined by the 
polarized quark, in particular the $u$-quark, distributions in the photon, 
which at large $x_{\gamma}$ are equal to their unpolarized counterparts as 
a result of the $Q^2$-evolution (see Fig.~2), rather {\em independently} of 
the hadronic input chosen. Thus the uncertainty coming from the polarized 
photon structure is minimal here and under control.
As becomes obvious from Fig.~4, the cross sections are fairly large over the 
whole range of $\bar{\eta}$ displayed and very sensitive to the shape 
{\em and} the size of $\Delta g$ with, unfortunately, not too sizeable 
asymmetries as compared to the statistical errors for ${\cal L}=100$/pb. 
A measurement of $\Delta g$ thus appears to be possible under the imposed 
conditions only if luminosities clearly exceeding $100$/pb can be reached. 
Fig.~5 displays the same results, but now for the resolved 
contribution with $x_{\gamma} \leq 0.75$ for the `maximally' saturated set 
(Figs.~5a,b) and the `minimally' saturated one (Figs.~5c,d). As expected, 
the results depend on the parton content of both the polarized photon and 
the proton, which implies that again the latter has to be known to some 
accuracy to allow for the extraction of some 
\begin{figure}[p]
\begin{center}
\vspace*{-1.8cm}
\hspace*{0.6cm}
\epsfig{file=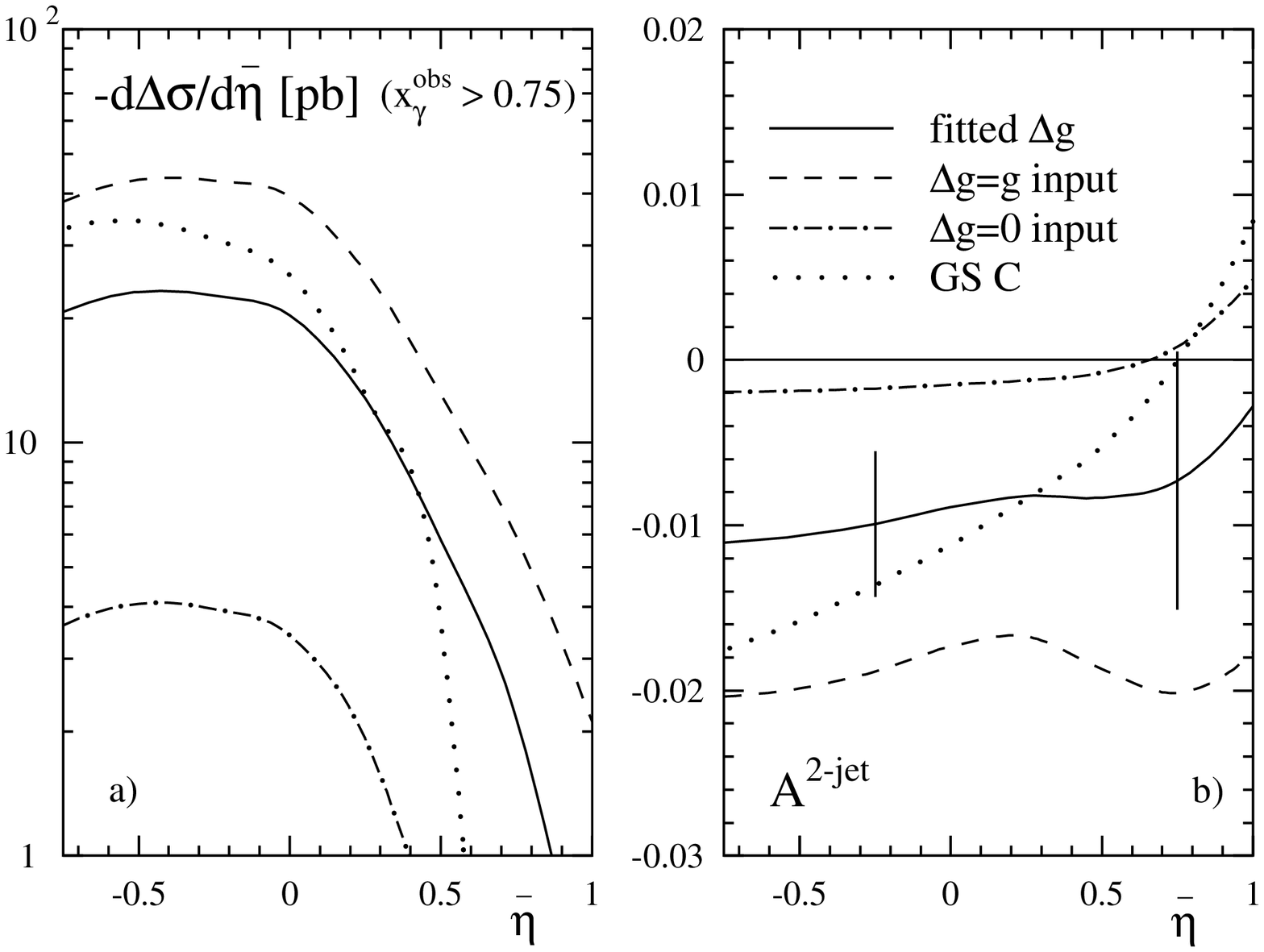,width=11cm}
\vspace*{-0.4cm}
\fcaption{\it {\bf a:} $\bar{\eta}$-dependence of the `direct' part 
($x_{\gamma}^{OBS}>0.75$) of the polarized two-jet photoproduction cross 
section in $ep$-collisions at HERA for the four different sets of polarized 
parton distributions of the proton. {\bf b:} Asymmetry corresponding to 
{\bf a}. The expected statistical errors indicated by the bars have been 
calculated according to (\ref{aerr}) and as explained in the text.}
\vspace*{-0.6cm}
%
\hspace*{1.3cm}
\epsfig{file=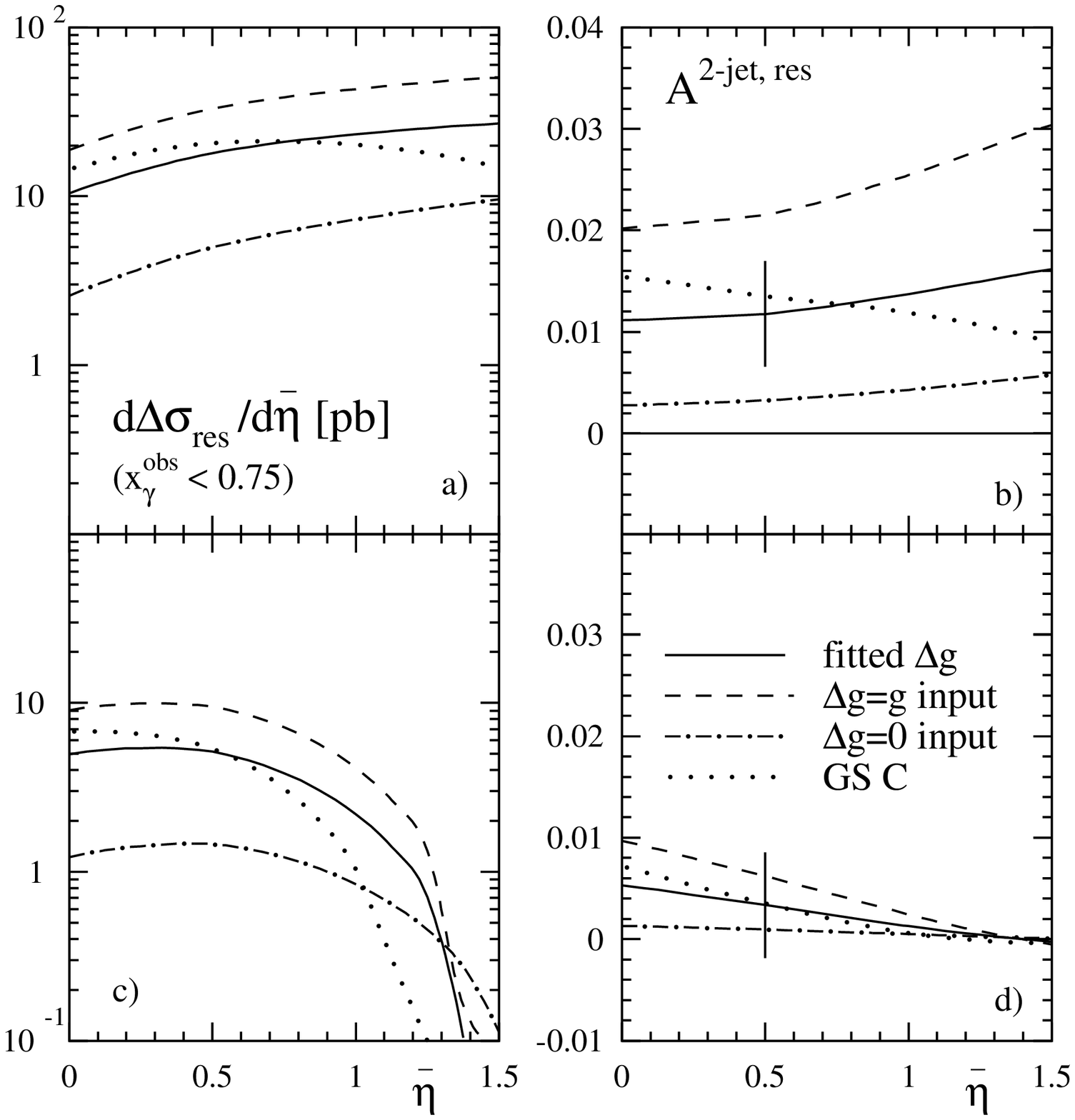,width=11cm}
\vspace*{-0.4cm}
\hspace*{0.5cm}
\fcaption{\it Same as Fig.~4, but for the resolved part of the cross section, 
defined by $x_{\gamma}^{OBS}\leq 0.75$ (see text). For {\bf a,b} the 
`maximally' saturated set of polarized photonic parton distributions has 
been used and for {\bf c,d} the `minimally' saturated one.}
\end{center}
\end{figure}
information on the polarized photon structure. 
We emphasize that the experimental finding of a non-vanishing 
asymmetry here would establish at least the definite existence of a resolved 
contribution to the polarized cross section. 

Let us finally discuss other photoproduction processes. From our results
for one-jet production in Fig.~3 it seems worthwhile to also consider the 
single-inclusive production of charged hadrons. At a first glance, this 
process appears less interesting than jet production, as the cross section 
for producing a definite hadron at a given $p_T$ will always be smaller 
than the one for a jet. On the other hand, in case of inclusive hadrons
one can obviously go experimentally to $p_T$ much smaller than the
$p_T^{min}=8$ GeV employed in our jet studies. Moreover, in the unpolarized
case single-inclusive hadron production was successfully studied 
experimentally at HERA prior to jets~\cite{ihh1,ihzeus}. The expression for 
the cross section for single-inclusive hadron production is similar to the 
one in (\ref{wqc}), but comprises an additional convolution with the 
function $D_c^h$ describing the fragmentation of particle $c$ into the
hadron $h$. For the $D_c^h$ we will use the LO fragmentation functions
of~\cite{bkk} which yield a good description of the unpolarized HERA inclusive
hadron data~\cite{ihh1,ihzeus}. Figs.~6a,b show our results for the sum 
of charged pions and kaons after integration over $p_T>3$ GeV, where all 
other parameters were chosen exactly as for Figs.~3a,b (since the sensitivity 
of the results to the polarized photon structure is qualitatively similar to 
the one-jet case we only consider the `maximally' saturated photon scenario 
here). One can see that the cross sections and
their asymmetries behave similarly in shape as the corresponding results
in Figs.~3a,b, but are somewhat smaller in magnitude. Nevertheless, the
expected statistical errors, calculated for the rather conservative choices 
$P_e * P_p=0.5$, ${\cal L}=100$/pb and $\epsilon=0.8$ in Eq.~(\ref{aerr}) 
and displayed in Fig.~6b, demonstrate that inclusive hadron photoproduction
remains a promising candidate.
\begin{figure}[h]
\begin{center}
\vspace*{-1cm}
\hspace*{0.5cm}
\epsfig{file=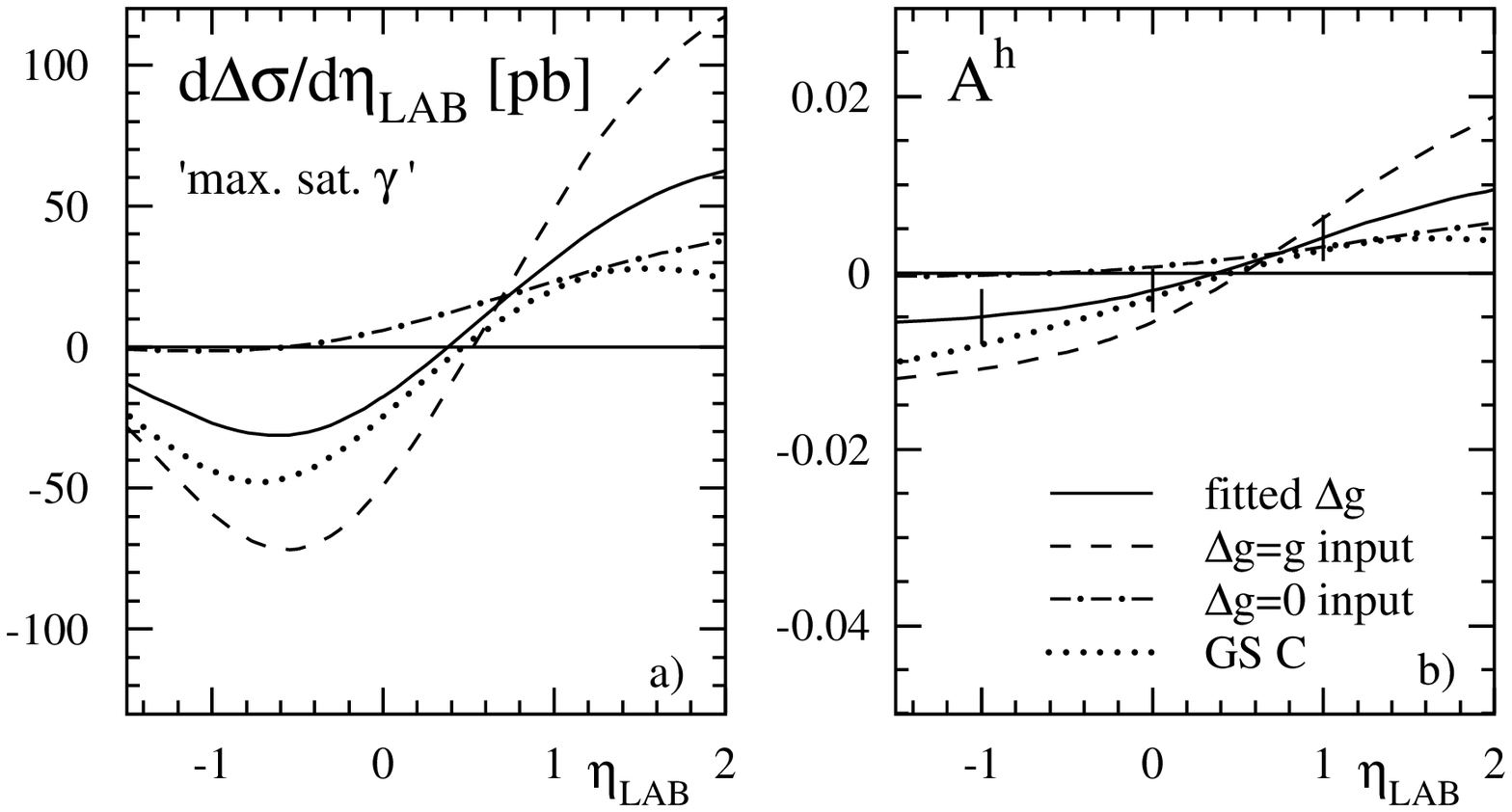,width=11.5cm}
\vspace*{-0.5cm}
\fcaption{\it {\bf a,b:} Same as Figs.~3a,b, but for the case of 
single-inclusive charged hadron production, integrated over $p_T>3$ GeV.}
\end{center}
\vspace*{-0.3cm}
\end{figure}

Another interesting process is polarized photoproduction of open 
charm~\cite{fr,wir}, which in the direct case should be mainly driven by 
the photon-gluon fusion subprocess $\gamma g \rightarrow c\bar{c}$
and in the resolved situation by $gg \rightarrow c\bar{c}$ and $q\bar{q} 
\rightarrow c\bar{c}$. However, a detailed study of this process for HERA 
conditions reveals~\cite{wir} that despite fairly large spin asymmetries for 
charm production the expected statistical errors for realistic luminosities 
and charm detection efficiencies will be too large for meaningful 
measurements.

To conclude, we have analyzed various photoproduction experiments in the 
context of a polarized $ep$-collider mode of HERA. We have found very
encouraging results for jet and inclusive-hadron production which 
look promising tools for a determination of the polarized gluon distribution 
of the proton and, possibly, might even allow access to the completely 
unknown parton content of a polarized photon. The proposed measurements 
will not be easy to do---but they seem a very interesting challenge for the 
future at HERA.

\end{document}